\documentstyle[prl,aps,twocolumn,epsfig]{revtex}
\begin{document}
\twocolumn[\hsize\textwidth\columnwidth\hsize\csname @twocolumnfalse\endcsname

\title{New Features of the Morphotropic Phase Boundary in the PbZr$_{1-x}$Ti$_x$O$_3$ System}
\author{B. Noheda, J. A. Gonzalo}
\address{Universidad Autonoma de Madrid, Cantoblanco, 28049-Madrid, Spain }
\author{A. Caballero, C. Moure}
\address{Instituto de Ceramica y Vidrio, CSIC, Arganda, 28500- Madrid, Spain}
\author{D.E. Cox, G. Shirane}
\address{Brookhaven National Laboratory, Upton, NY 11973-5000, USA}

\date{July 20, 1999}

\maketitle

\begin{abstract}
Recently a new monoclinic phase in the PbZr$_{1-x}$Ti$_x$O$_3$ ceramic
system has been reported by Noheda et al.\cite{Noheda} for the composition
x= 0.48. In the present work, samples with Ti contents of x= 0.47 and 0.50,
which are both tetragonal below their Curie points, have been investigated.
In the sample with x= 0.50, the tetragonal phase was found to transform to a
monoclinic phase at about 200 K as the temperature was lowered. The sample
with x= 0.47 showed a complicated region of phase coexistence between $\sim$
440-320 K, becoming rhombohedral at around 300 K. No further symmetry change
was found down to 20 K. Dielectric measurements for these two samples are
also reported. On the basis of these results, a preliminary phase diagram is
presented. Optimum compositional homogeneity is needed to properly
characterize the new monoclinic region.
\end{abstract}
\vskip1pc]

%\begin{multicols}{1}
\narrowtext

%\preprint{cond-mat/000-ms}

\section{Introduction}

\label{sec:level1}

The basic features of the PbZr$_{1-x}$Ti$_x$O$_3$ (PZT) phase diagram were
determined in the 1950's\cite{Shirane}. The ceramic PZT system has the cubic
perovskite ABO$_3$ structure at high temperatures. On lowering the
temperature the materials undergo a phase transition to a ferroelectric
phase for all compositions except those close to pure PbZrO$_3$, where they
become antiferroelectric. The ferroelectric region is divided in two phases
with different symmetries by a morphotropic phase boundary (MPB), nearly
vertical in temperature, occurring at a composition close to x= 0.47. The
Ti-rich region has tetragonal symmetry ($F_T$, space group P4mm) and the
Zr-rich region has rhombohedral symmetry ($F_R$). The latter is divided into
high-temperature ($F_{R(HT)}$, space group R3c) and low-temperature ($%
F_{R(LT)}$, space group R3m) zones respectively \cite{Barnett}\cite{Michel}%
\cite{Glazer}.

Most of the studies on PZT have been performed for compositions around the
MPB, motivated both by the interesting physical properties and the
technologically-useful applications, such as high electromechanical coupling
factors and permittivities, exhibited by PZT at this boundary \cite{Jaffe}.
Due to compositional fluctuations, the MPB often appears as an ill-defined
region of phase coexistence, instead of a well-defined boundary, whose size
depends of the processing conditions \cite{Ari-Gur}\cite{Kakewaga}\cite
{Fernandes}. This fact has for many years hindered a detailed interpretation
of the nature of the $F_R-F_T$ phase transition and the MPB itself. In order
to explore the MPB in more detail, we have embarked upon a systematic
structural study using high-resolution synchrotron x-ray powder diffraction
techniques and dielectric measurements to characterize the samples.

A recent unexpected result obtained from high-resolution x-ray measurements
on a sample of high compositional homogeneity with x=0.48, was the discovery
of a ferroelectric phase with monoclinic symmetry below approximately 250 K 
\cite{Noheda}. The monoclinic unit cell is such that $a_m$ and $b_m$ lie
along the tetragonal [$\overline{1}\overline{1}$0] and [1$\overline{1}$0]
directions ($a_m\approx b_m\approx a_t\surd 2$), and $c_m$ is close to the
[001] axis ($c_m\approx c_t$). The temperature dependence of the monoclinic
angle $\beta$, the angle between $a_m$ and $c_m$, corresponds to the
evolution of the order parameter for the tetragonal-monoclinic ($F_T-F_M$)
phase transition. Two other compositions prepared under slightly different
conditions with x = 0.47 and 0.50 were also studied at that time, and in the
present paper we report results for these materials and propose a
preliminary modification of the PZT phase diagram around the MPB.

\section{Experimental}

\label{sec:level2}

Two different compositions of PZT with Ti contents of 0.47 and 0.50 were
prepared by a solid-state reaction from PbO$_2$, ZrO$_2$ and Nb-free TiO$_2$
with chemical purities better than 99.9\%. The mixed powders were calcined
at 790$^o$C, remilled, isostatically pressed at 200 Mpa, and sintered at 1200%
$^o$C for 2h, with heating and cooling rates of 3$^o$C/min. To minimize the
volatilization of lead oxide, alumina crucibles with tightly-fitting covers
were used, and a mixture of PbZrO$_3$+5 wt\% ZrO$_2$ was used as a lead
source in the crucible. The densities measured by the liquid displacement
method were $\geq$98\% of the theoretical values.

High-resolution synchrotron x-ray powder diffraction measurements were made
at beam line X7A at the Brookhaven National Synchrotron Light Source. A
Ge(111) double-crystal monochromator was used in combination with a Ge(220)
analyser, with wavelengths of 0.6896 \AA~for x= 0.47 and 0.7995 \AA~for x=
0.50. In this configuration, the instrumental resolution, $\Delta 2\theta$,
is slightly better than 0.01$^o$ in the $2\theta$ region 0-30$^o$, an
order-of-magnitude better than that of a conventional laboratory instrument.
The pellets were mounted in symmetric reflection geometry and scans made
over selected peaks in the low-angle region of the pattern. Since lead is a
strong absorber, the penetration depth below the surface of the pellet at $%
2\theta=20^o$ is only about 2 $\mu$m. Measurements were made at various
temperatures between 20-790 K for x= 0.47, and from 20-300 K for x= 0.50.
\begin{figure}[h]
\epsfig{width=0.80 \linewidth,figure=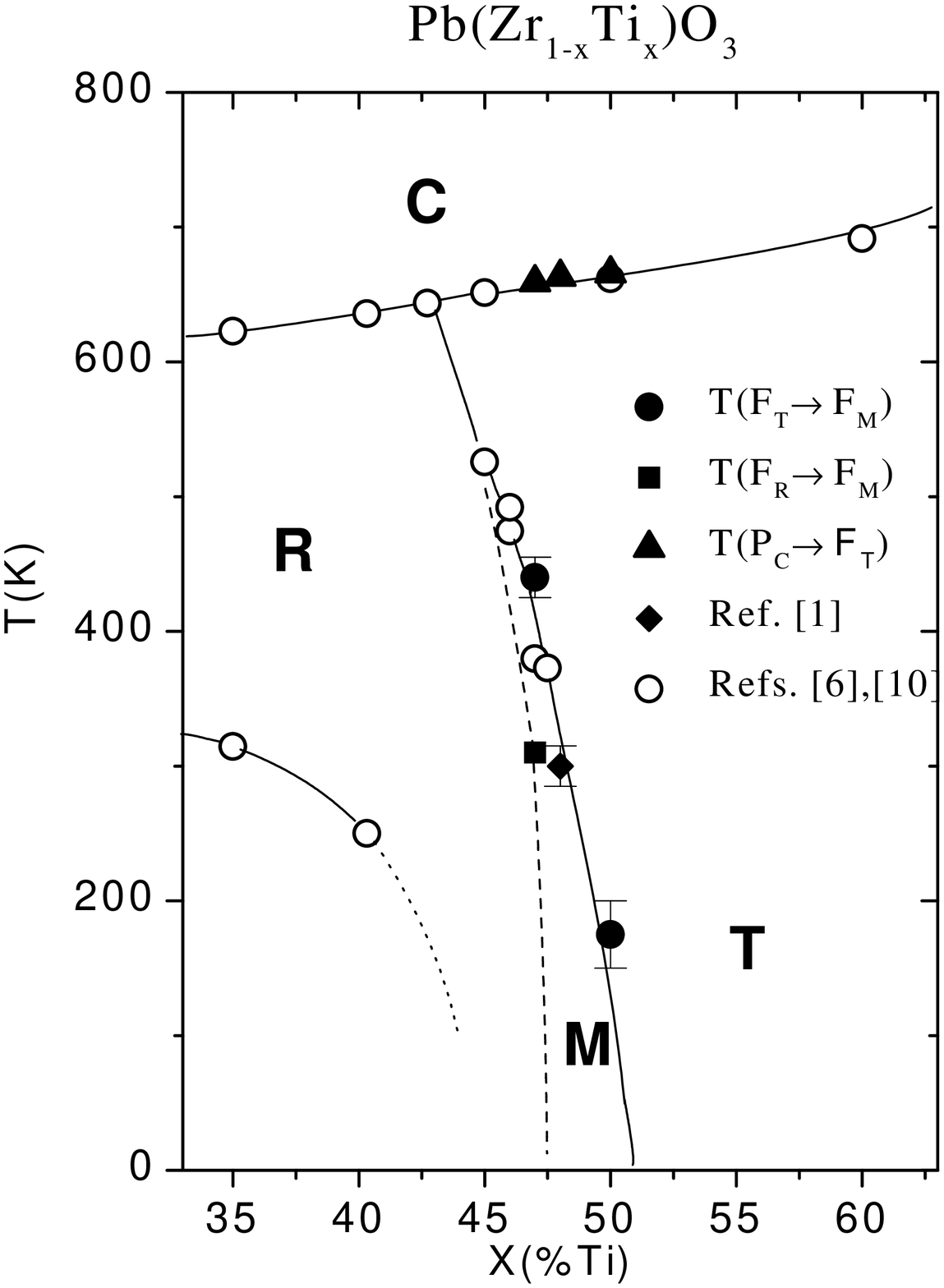} %vbox{\vspace{1.9truecm}} 
%\hbox{\hspace{0.1truecm}\special{illustration Fig1FH.eps scaled 920}} 
\vspace{0.0truecm}
\caption{Preliminary modification of the PZT phase diagram around its MPB
obtained from the results of this work. Data in refs.[6] (p.136) and [10]
(x=0.40) are plotted as open cicles. Data in ref.[1] (x= 0.48) are also
included.}
\label{fig1}
\end{figure}
Measurements above room temperature were performed with the pellet mounted
on a flat BN sample holder inside a wire-wound BN tube furnace. The accuracy
of the temperature was estimated to be within 5 K, and the temperature
stability was $\sim$2 K. For measurements below room temperature, the sample
was mounted on a flat Cu sample holder in a closed-cycle He cryostat. In
this case, the estimated accuracy of the temperature was 1 K, with a
stability of $\sim$0.1 K. The angular regions scanned were chosen so as to
cover the pseudo-cubic (100), (110), (111), (200), (220) and (222)
reflections, with a 2$\theta $ step interval of 0.005 or 0.01$^o$ depending
on the peak widths. Dielectric measurements were performed with a precision
LCR meter (Hewlett Packard-4284A) varying temperature at a constant rate of
0.5 K/min with a temperature accuracy better than 0.1 K.

\section{Results and Discussion}

\label{sec:level3}

Based on the analysis of the diffraction data for the PZT compositions
studied in this work (x= 0.47 and 0.50) and that previously reported for x=
0.48 \cite{Noheda}, we propose a modification of the PZT phase diagram \cite
{Jaffe} around its MPB, which includes the new monoclinic phase as shown in
Fig. 1, where temperatures below 300 K are also shown. As can be seen, the
MPB in Jaffe's phase diagram corresponds to the phase boundary between the $%
F_T$ and the $F_M$ phases, but the $F_M-F_R$ phase boundary is still not
well defined.

\begin{figure}[h]
\epsfig{width=0.85 \linewidth,figure=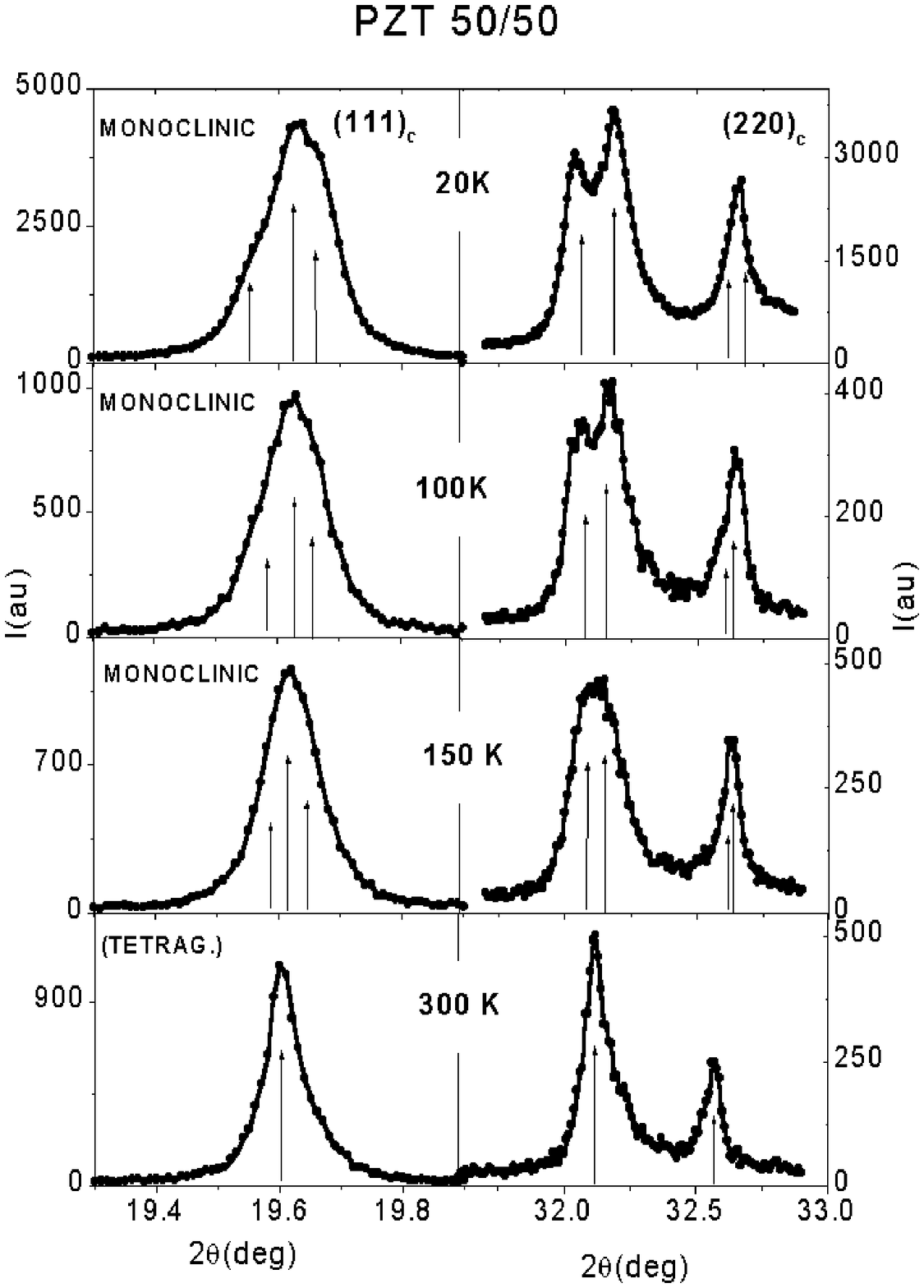} %vbox{\vspace{1.9truecm}} 
%\hbox{\hspace{0.1truecm}\special{illustration Fig2FH.eps scaled 920}} 
\vspace{0.0truecm}
\caption{$(111)_c$ and $(220)_c$ pseudo-cubic reflections at different
temperatures for PbZr$_{0.50}$Ti$_{0.50}$O$_3$.}
\label{fig2}
\end{figure}

Fig. 2 shows that the monoclinic phase is found to exist for x= 0.50 below
room temperature. From peak fits based on a pseudo-Voigt peak function, at
20 K the (111)$_c$ pseudocubic reflection is found to consist of three
different peaks corresponding to the monoclinic ($\overline{2}$01), (021)and
(201) reflections, while the pseudocubic (220)$_c$ is split into four peaks
corresponding to the ($\overline{2}$22),(222), (400) and (040) monoclinic
reflections. The last two are fairly close to each other, indicating that
the difference between $a_m$ and $b_m$ is quite small. As the temperature
increases the monoclinic splitting becomes less evident. For T= 150 K, (400)
and (040) at 2$\theta \approx 32.62^{\circ}$ cannot be resolved, showing
that $a_m\approx b_m$. For T%
%TCIMACRO{\TEXTsymbol{>} }
%BeginExpansion
\mbox{$>$}
%EndExpansion
200 K, the monoclinic features, if any, cannot be detected, and the observed
reflections can be indexed on the basis of a tetragonal unit cell. The
evolution of the lattice parameters for temperatures below 300 K is shown in
Fig. 3.

\begin{figure}[h]
\epsfig{width=0.70 \linewidth,figure=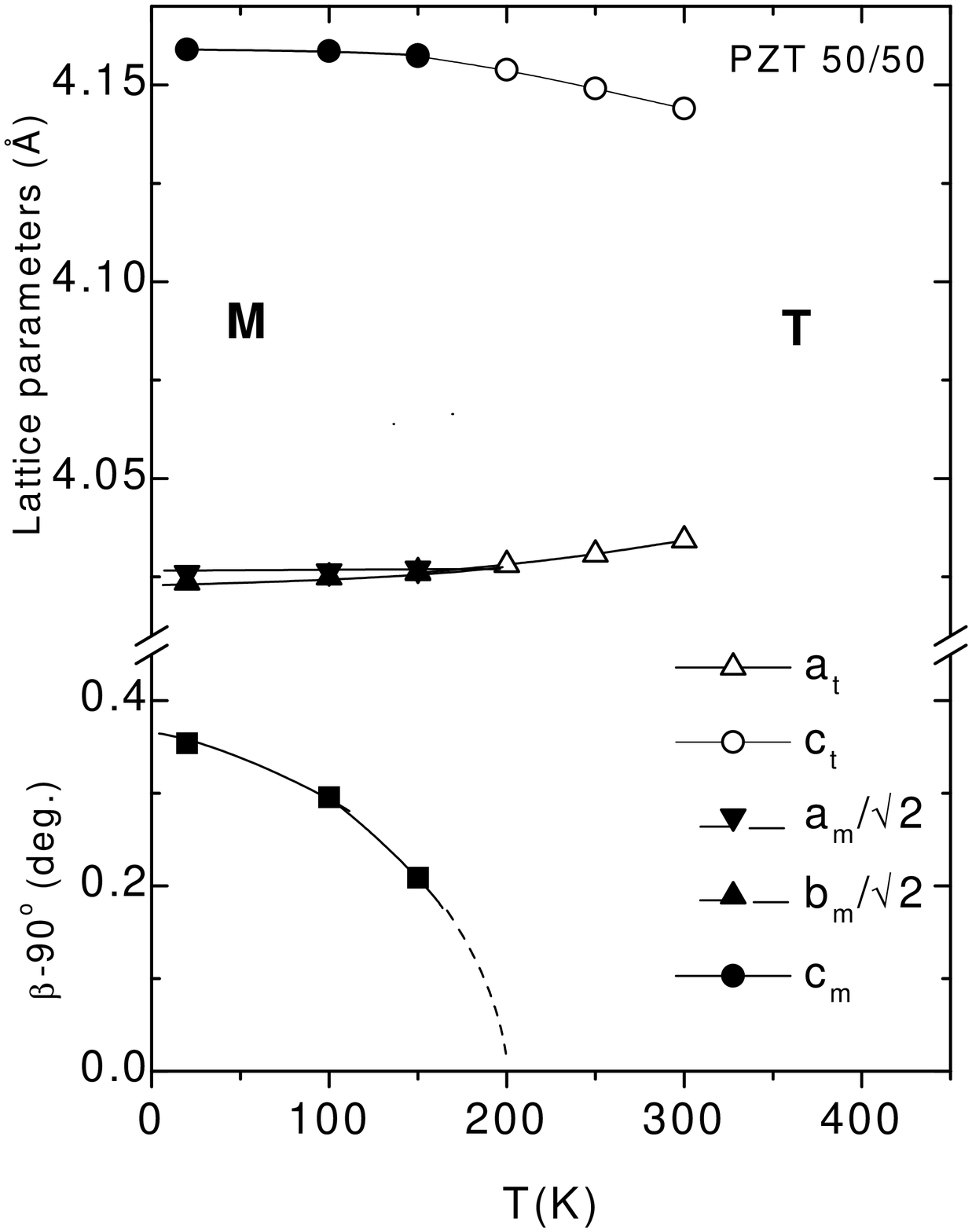} %vbox{\vspace{1.9truecm}} 
%\hbox{\hspace{0.1truecm}\special{illustration Fig3FH.eps scaled 920}} 
\vspace{0.0truecm}
\caption{Lattice parameters vs. temperature for x= 0.50 for the monoclinic ($%
a_m$, $b_m$ , $c_m$, $\beta$) and tetragonal ($a_t$, $c_t$) phases.}
\label{fig3}
\end{figure}

The PZT composition with x= 0.47 was found to be rhombohedral from 20-300 K.
Fig. 4 shows the temperature evolution of the (111)$_c$ and (200)$_c$
pseudo-cubic reflections between 300-787 K. At 300 K (111)$_c$ is split into
rhombohedral (111) and (11$\overline{1}$) peaks, while (200)$_c$ remains a
single peak, corresponding to rhombohedral (200). For 300 K%
%TCIMACRO{\TEXTsymbol{<} }
%BeginExpansion
\mbox{$<$}
%EndExpansion
T%
%TCIMACRO{\TEXTsymbol{<} }
%BeginExpansion
\mbox{$<$}
%EndExpansion
440 K there is a region where the peak profiles broaden in a complex way,
suggesting the gradual evolution of a second phase which is difficult to
characterize (see T= 372 and 425 K in Fig. 4). 
\begin{figure}[h]
\epsfig{width=0.85 \linewidth,figure=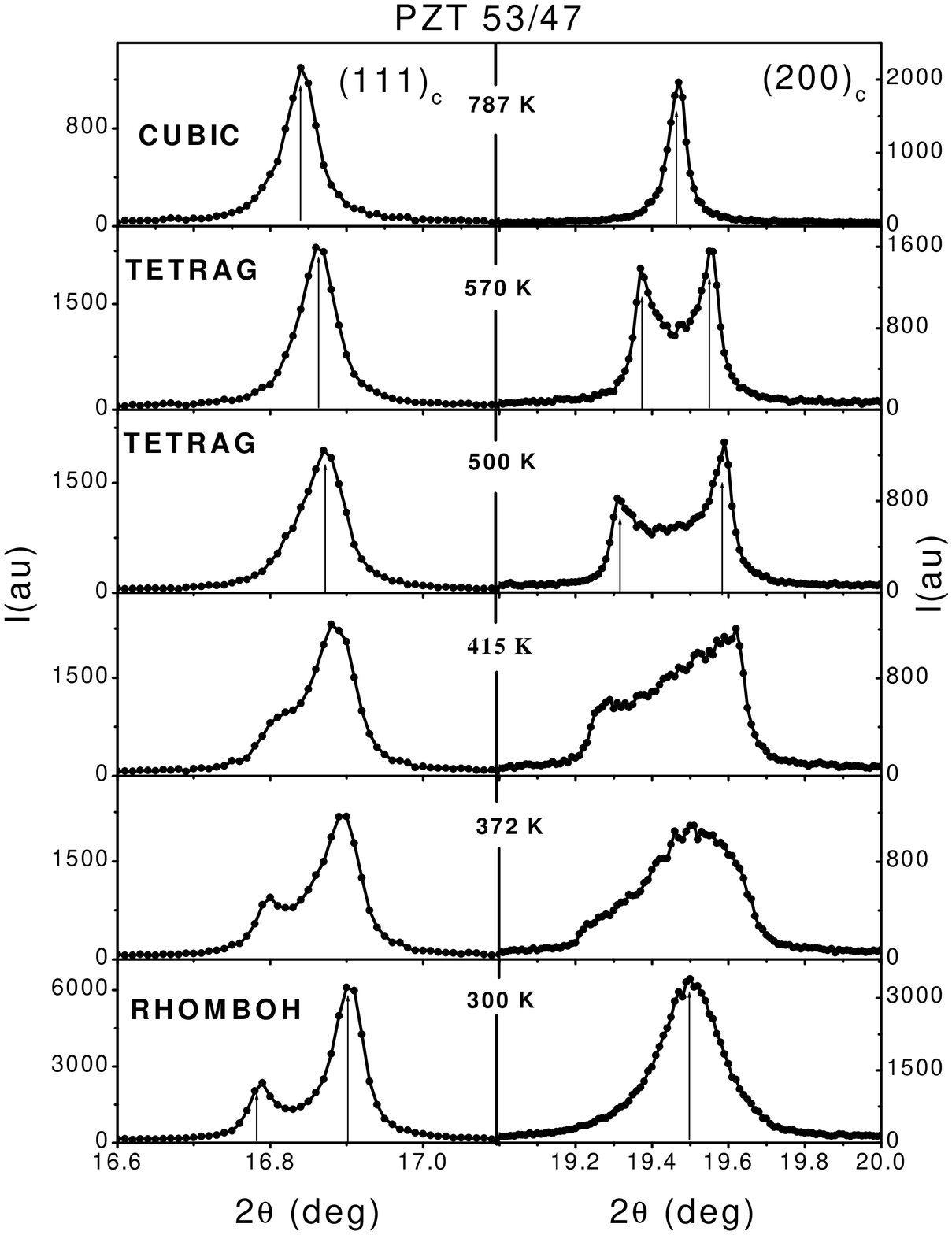} %vbox{\vspace{1.9truecm}} 
%\hbox{\hspace{0.1truecm}\special{illustration Fig4FH.eps scaled 920}} 
\vspace{0.0truecm}
\caption{$(111)_c$ and $(200)_c$ pseudo-cubic reflections at different
temperatures for PbZr$_{0.53}$Ti$_{0.47}$O$_3$. Note the improved resolution
compared to the laboratory x-ray data reported by Mishra et al.}
\label{fig4}
\end{figure}
This could simply reflect the coexistence of rhombohedral and tetragonal
phases accompanied by a considerable amount of internal strain, but it is
certainly not possible to rule out the formation of small local regions of
the monoclinic phase. For T above $\sim$450 K, the tetragonal phase can be
clearly identified although there is still some residual diffuse scattering
in the vicinity of rhombohedral (200). Finally, at T$\approx 665$ K, the
cubic phase appears. The temperature evolution of the lattice parameters is
shown in Fig. 5. We note that the peak profiles in the cubic region are
about twice as broad as those found for the x = 0.48 sample \cite{Noheda},
indicative of a smaller crystallite size and wider range of compositonal
inhomogeneity. This is probably associated with the slightly different
sintering temperatures used for the preparation of the samples, 1200 and 1250%
$^o$C respectively.
\begin{figure}[h]
\epsfig{width=0.95 \linewidth,figure=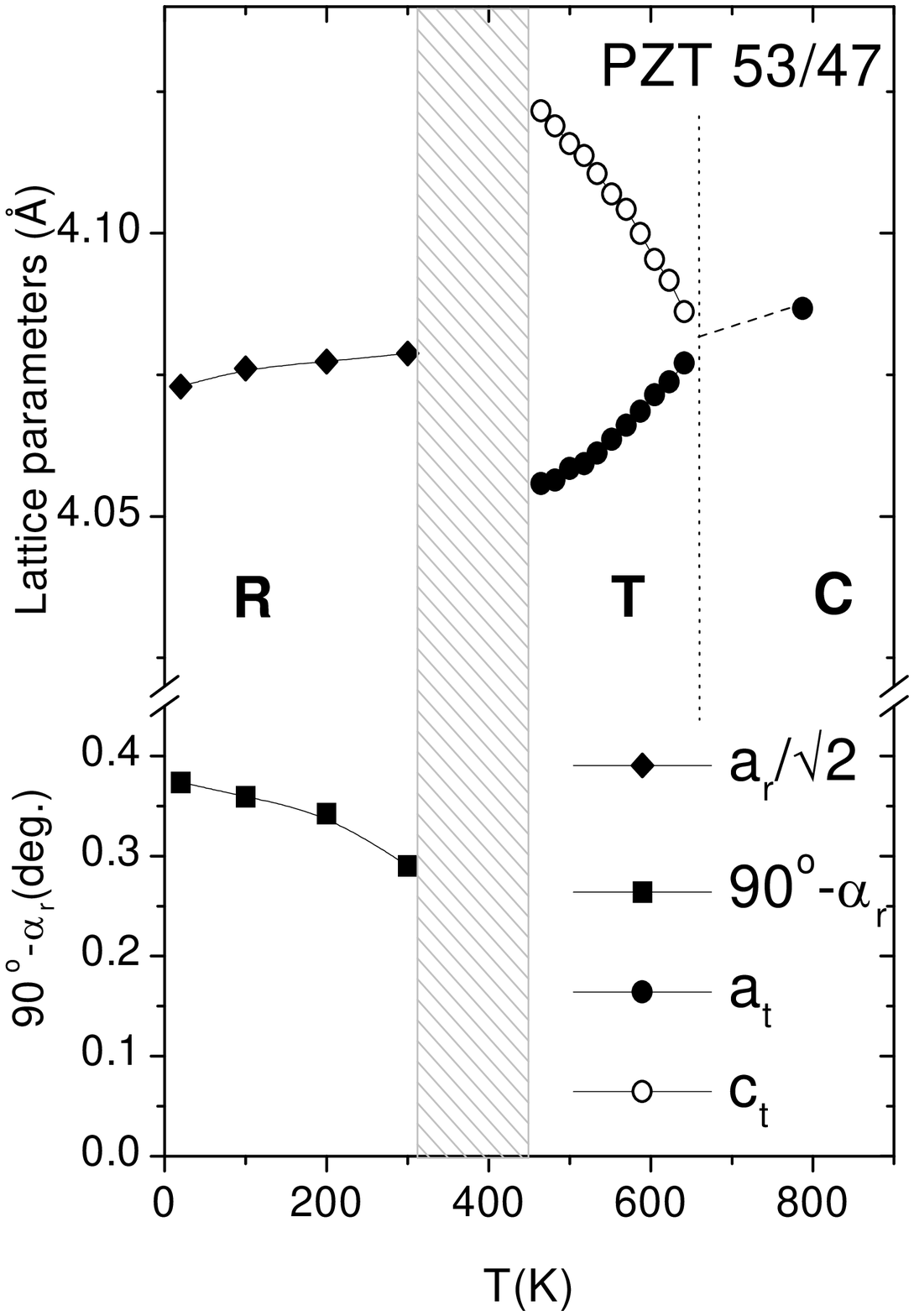} %vbox{\vspace{1.9truecm}} 
%\hbox{\hspace{0.1truecm}\special{illustration Fig5FH.eps scaled 920}} 
\vspace{0.0truecm}
\caption{Lattice parameters vs. temperature for x= 0.47 for the rhombohedral
($a_r, \alpha_r $) and tetragonal ($a_t, c_t$) phases. The complex region of
phase coexistence is shown shaded in the plot.}
\label{fig5}
\end{figure}
The dielectric permittivity, $\varepsilon $, was measured along the axis of
the pellets at 1, 10 and 100 kHz for both compositions as the temperature
was raised from 140 K. $\varepsilon ^{-1}$ is plotted in Fig. 6 for x= 0.50
in the interval 140 K 
%TCIMACRO{\TEXTsymbol{<} }
%BeginExpansion
\mbox{$<$}
%EndExpansion
T%
%TCIMACRO{\TEXTsymbol{<} }
%BeginExpansion
\mbox{$<$}
%EndExpansion
350 K. Two changes of slope are clearly observed as temperature values
decrease, the first one at ~$\sim $310 K, and the second at ~$\sim $190 K.
It is possible that the first of these two anomalies could correspond to the
onset of a local monoclinic distortion even though we were unable to resolve
any monoclinic splitting ($a_m=b_m$). The anomaly at T$\approx $ 190 K would
then correspond to the onset of the long-range distortion below which $%
a_m\neq b_m$.

The inverse of the dielectric permittivity with increasing temperature is
plotted in Fig. 7 for x= 0.47. In the region below the cubic-tetragonal
transition at 640 K, there is an increase in slope in the region around 440
K as indicated by the broken lines, while at lower temperatures the slope
continues to increase, but without any sharp discontinuities indicative of a
well-defined transition. The results are generally consistent with the
diffraction evidence for phase coexistence and the possible existence of a
monoclinic phase. The permittivity data are in good agreement with those
recently reported by Zhang et al.\cite{Zhang}, who also found thermal
hysteresis effects between 300-400 K which they interpreted as the
coexistence of rhombohedral and tetragonal phases. A R-T coexistence region
between 473-533 K was also inferred by Mishra et al.\cite{Mishra} for a
sample with x=0.535 on the basis of planar coupling coefficient measurements
and laboratory x-ray data.

\begin{figure}[h]
\epsfig{width=0.85 \linewidth,figure=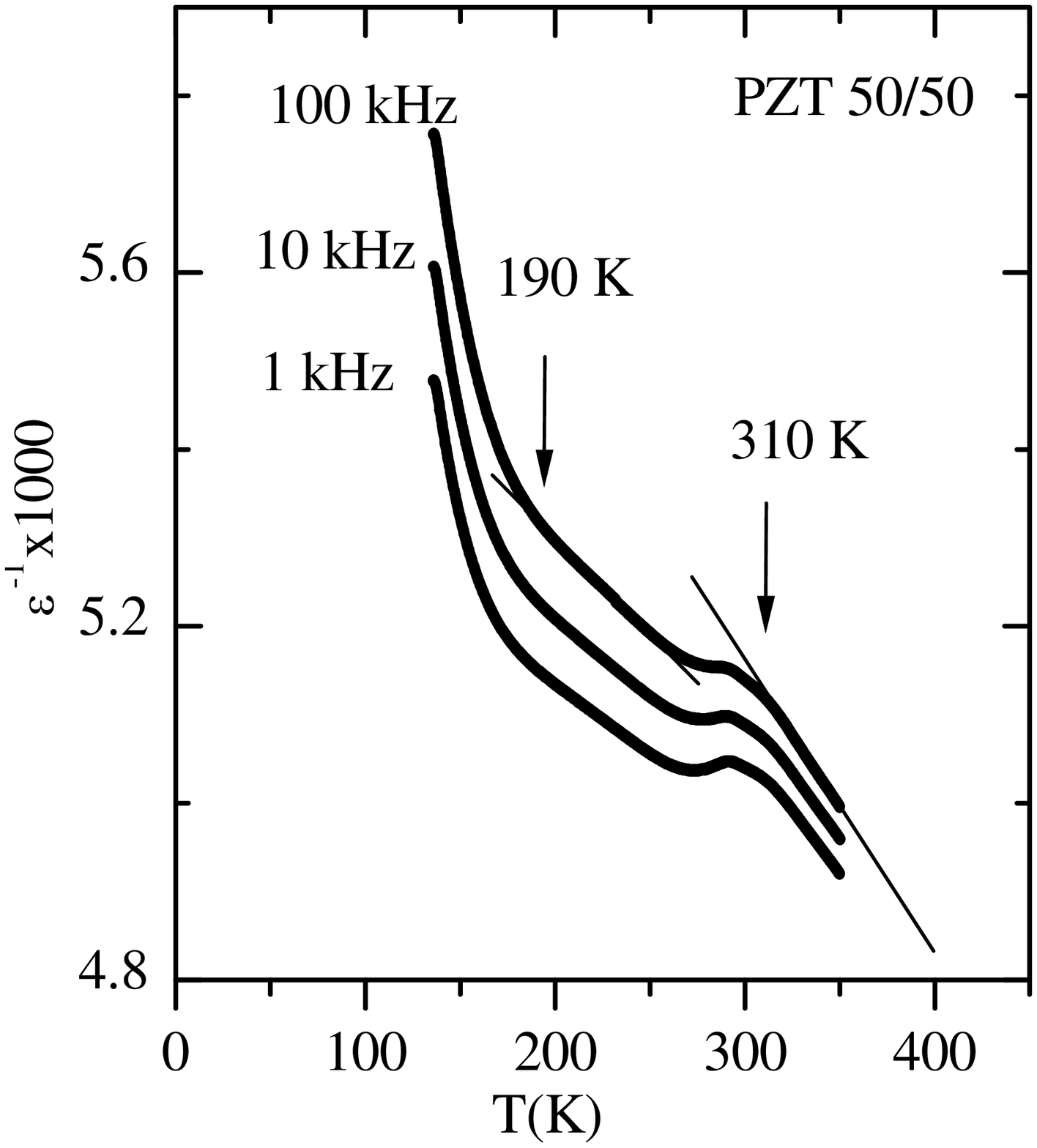} %vbox{\vspace{1.9truecm}} 
%\hbox{\hspace{0.1truecm}\special{illustration Fig6FH.eps scaled 920}} 
\vspace{0.0truecm}
\caption{Inverse of the dielectric permittivity vs. temperature for three
different frequencies for PbZr$_{0.50}$Ti$_{0.50}$O$_3$.}
\label{fig6}
\end{figure}

This work clearly demonstrates the need for both excellent compositional
homogeneity and high instrumental resolution for the determination of the
features of the PZT phase diagram around the MPB. Further work along these
lines is in progress.

\section{Acknowledgments}

We thank L.E. Cross and R. Guo for their stimulating discussions. Support by
NATO (R.C.G.970037), the Spanish CICyT (PB96-0037) and the U.S. Department
of Energy (contract No. DE-AC02-98CH10886) is also acknowledged.

\begin{figure}[h]
\epsfig{width=0.75 \linewidth,figure=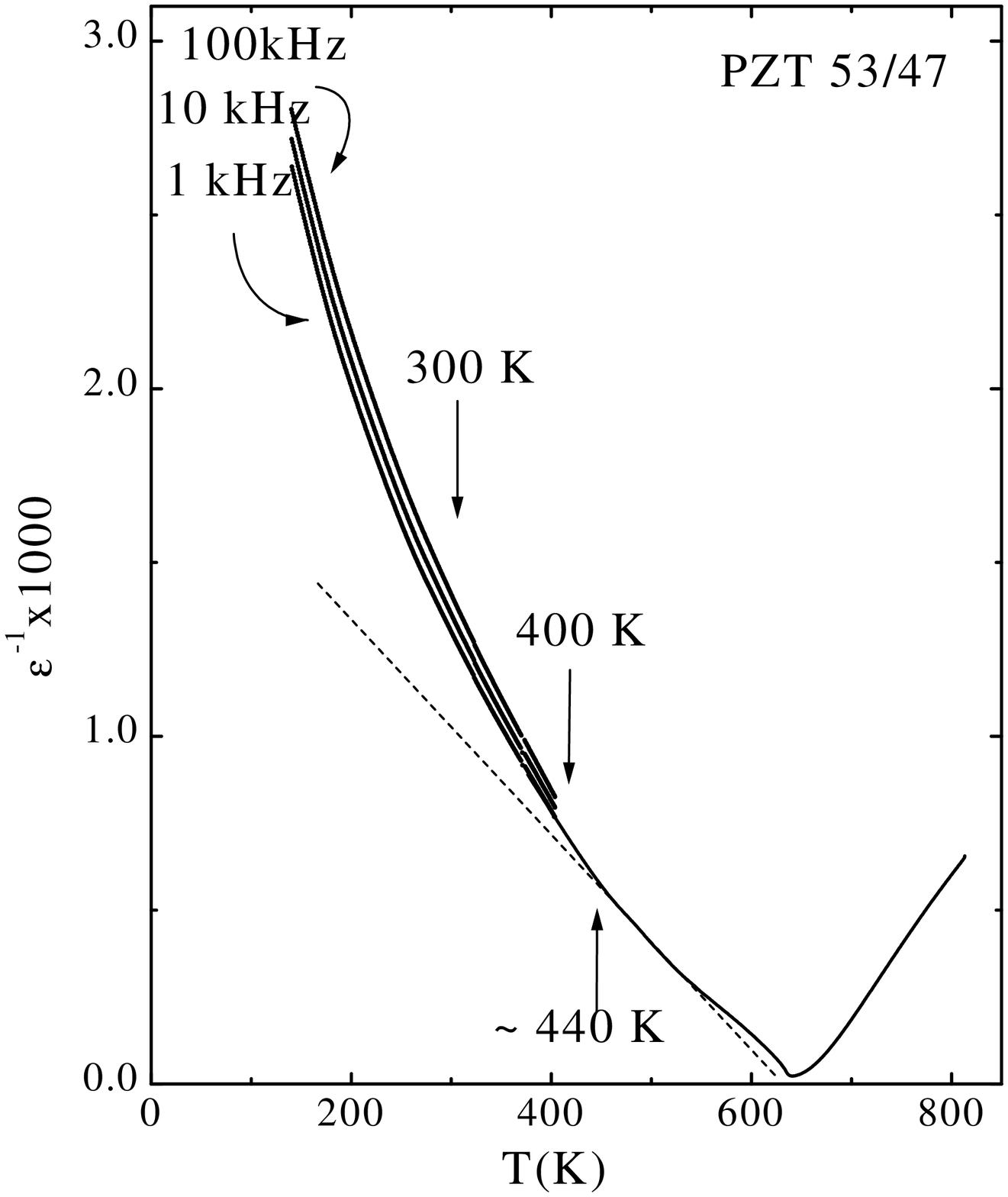} %vbox{\vspace{1.9truecm}} 
%\hbox{\hspace{0.1truecm}\special{illustration Fig7FH.eps scaled 920}} 
\vspace{0.0truecm}
\caption{Inverse of the dielectric permittivity vs. temperature for three
different frequencies for PbZr$_{0.53}$Ti$_{0.47}$O$_3$. At T$>400$ K only
data at 1 kHz are shown for clarity. Note the change of slope at $\approx $%
440 K shown by the broken line, and the two arrows at 300 and 400 K,
which correspond to the region of thermal hysteresis reported by Zhang et
al. }
\label{fig7}
\end{figure}


\begin{references}
\bibitem{Noheda}  B. Noheda, D.E. Cox, G. Shirane, J.A. Gonzalo, S-E. Park,
L.E. Cross. Appl. Phys. Lett. 74(14), 2059 (1999). $\prec $http://xxx.lanl.gov/abs/cond-mat/9903007$\succ $

\bibitem{Shirane}  G. Shirane and K. Suzuki. J. Phys. Soc. Japan 7, 333
(1952). E. Sawaguchi. J. Phys. Soc. Japan 8, 615 (1953).

\bibitem{Barnett}  H. Barnett, J. Appl. Phys. 33, 1606, (1962).

\bibitem{Michel}  C. Michel, J.N. Moreau, G.D. Achenbach, R. Gerson, W.J.
James. Solid State Commun., 7, 865 (1969).

\bibitem{Glazer}  A. M. Glazer and S. A. Mabud. Acta Cryst. B34, 1060(1978).

\bibitem{Jaffe}  B. Jaffe, W.R. Cook, and H. Jaffe, Piezoelectric Ceramics,
Academic Press, London (1971).

\bibitem{Ari-Gur}  P. Ari-Gur and L. Benguigui. Sol. Stat. Commun. 15 , 1077
(1974).

\bibitem{Kakewaga}  K. Kakewaga, O. Matsunaga, T. Kato and Y. Sasaki. J.
Amer. Ceram. Soc. 78, 1071 (1995).

\bibitem{Fernandes}  J.C. Fernandes, D.A. Hall, M.R.Cockburn and
G.N.Greaves. Nucl. Instrum. Meth. B97, 137 (1995).

\bibitem{Amin}  A. Amin, R.E. Newnham, L. E. Cross, D. E. Cox. J. Solid
State Chem. 37, 248 (1981).

\bibitem{Zhang}  S. Zhang, X. Dong, S. Kojima. Jpn. J. Appl. Phys. 36,
2994(1997).

\bibitem{Mishra}  S. K. Mishra, D. Pandey. Appl. Phys. Lett. 69, 1707
(1996)
\end{references}
\end{document}